\documentclass[prl,twocolumn,aps,showpacs,floatfix]{revtex4}
\usepackage{graphicx}
\usepackage{amssymb,amsmath}
\usepackage{textcomp}

\newcommand{\br}    {{\bf r}}
\newcommand{\bx}    {{\bf x}}
\newcommand{\bq}    {{\bf q}}
\newcommand{\bp}    {{\bf p}}

\newcommand{\Schro} {Schr\"odinger }

\begin{document}

\title{Correlated Electron Transport}

\author{P. Delaney and J.C. Greer} 
\affiliation{NMRC, University College, Prospect Row, Cork, Ireland}
\email{paul.delaney@nmrc.ie, jim.greer@nmrc.ie}
\homepage{http://www.nmrc.ie/research/computational-modelling-group/index.html}
\date{\today}

\begin{abstract}
Theoretical and experimental values to date for the resistances of single molecules
commonly disagree by orders of magnitude. 
By reformulating the transport problem using boundary conditions suitable for correlated many-electron systems,
we approach electron transport across molecules from a new standpoint. 
Application of our correlated formalism to benzene-dithiol gives current-voltage characteristics
close to experimental observations.
The method can solve the open system quantum many-body problem accurately,
treats spin exactly and is valid beyond the linear response regime. 
\end{abstract}

\maketitle


Substantial interest in recent years has been applied to extend the machinery
of electronic structure theory, largely developed for study of  
closed or periodic systems, to open quantum systems.
A primary motivation for this effort is the study of electronic transport for nanoscale 
systems, and specifically the study of molecular electronics~\cite{RZM97,JGA00,NiR03}.
Recent approaches have introduced density functional theory (DFT)
into scattering~\cite{EmK98,DPL00} or Green's function calculations of transport~\cite{DeS01,DGD01,STB03,Krs03}.
The calculated single-electron transmission coefficients ${\rm T}$ are then used to
predict a molecular conductance using the Landauer-B\"uttiker formula $\sigma = {\rm T G}_0$,
where ${\rm G}_0 = 2 e^2/h$ is the conductance quantum~\cite{LanBue}.
Notwithstanding this effort, 
the currents flowing through molecular scale systems are consistently predicted to
be several orders of magnitude larger than experimentally reported values~\cite{RZM97,NiR03,Lee03}. 
By formulating scattering boundary conditions appropriate to many-body wavefunctions,
we are able to present correlated transport calculations which show
that an accurate description of electron correlation is essential for single molecule transport.
We find current magnitudes close to experiment for the prototypical molecular electronics system,
benzene-dithiol bonded between two gold contacts.

Central to the study of electronic transport is the concept of two reservoirs 
locally in equilibrium, but driven out of equilibrium with respect to one another. 
The corresponding electrochemical
potential imbalance is interpreted as the voltage driving a current across a 
device in contact with both reservoirs. 
Typically, scattering boundary conditions are introduced to define the properties of the electron reservoirs:
one reservoir (the left) is associated with the emission of incoming, right moving particles
with a fixed equilibrium energy distribution, but capable of absorbing outgoing, left moving particles
at any energy; the converse is true for the right reservoir.
Implementing scattering boundary conditions for a system of fermions within a 
single particle approximation is straightforward. 
Fermi-Dirac statistics are applied to the left and right reservoir incoming wavefunctions $\psi_i^{L,R}(\br)$ 
and a voltage is introduced as the difference in chemical potential between 
the left and right reservoir distribution functions $n_F(\epsilon_i \pm eV/2)$ inducing a net current flow.
When this scheme is applied with commonly used DFT computations, several formal and technical problems arise. 
For example, the exchange-correlation functional is current dependent \cite{ViR87},
a fact not explored to date within DFT transport studies.
Even ignoring this current dependence, a choice for approximating the exchange-correlation functional must be made.
Recently, it has been pointed out that the currents theoretically predicted with DFT methods
can vary by over an order of magnitude simply due to the choice of exchange-correlation functional \cite{Krs03}.
Another drawback to DFT transport implementations is the assumption that the Kohn-Sham orbitals $\psi_i$ are
physical single particle states and give the correct transmission coefficients. 
Use of the Landauer-B\"uttiker formula beyond the linear response regime
also introduces a level of approximation which is difficult to control.

In our approach to the electronic transport problem, we avoid these
issues by working directly with many-body wavefunctions $\Psi(\br_1 s_1, \ldots , \br_N s_N)$
and the exact molecular electronic Hamiltonian $\hat{H}$.
By discovering that a form of the single-electron scattering boundary conditions~\cite{Fre90} can be generalized
to many-body wavefunctions, we can apply many-body methods to the electronic structure of open systems. 
Here we use the configuration interaction (CI) formalism which can give accurate solutions to quantum many-body 
problems with correct spin eigenfunctions and an equal treatment of ground and excited electronic states;
our approach is equally applicable to any other correlated wavefunction method. 
In CI we expand the N-electron wavefunction $\Psi$ in terms of spin-projected Slater determinants
\begin{eqnarray}
\Psi & = & c_1 \Psi_1 + c_2 \Psi_2 + \ldots c_L \Psi_L \label{eq:CI_wavefunction}
\end{eqnarray}
with expansion coefficients $c_\mu$. 
We use the exact non-relativistic electronic Hamiltonian
\begin{equation}
\hat{{\rm H}} = - \sum_{n=1,N} \frac{\hbar^2}{2m} \nabla_n^2 + 
\sum_{n=1,N} {\rm U}( \br_n ) + 
\sum_{ n<m } \frac{e^2}{| \br_n - \br_m |} \label{eq:Hdef}
\end{equation}
where ${\rm U}( \br )$ is the attractive potential energy of an electron in the Coulomb field of the atomic nuclei.
We then diagonalize the matrix ${\rm H}_{\mu \nu}$ of the Hamiltonian in the basis $\Psi_\mu$
to find the many-body eigenstates and energies.

Working directly with the N-particle wavefunction $\Psi$ removes a direct physical interpretation
of single-electron wavefunctions $\psi_i(\br)$ and eigenvalues $\epsilon_i$ from the description
of many-electron systems.
Therefore, generalization of the conventional single-electron scattering boundary conditions is not possible,
and a formulation of the transport problem does not exist for correlated many-electron systems. 
We resolve this issue by formulating the scattering boundary conditions in terms of the Wigner function $f(q,p)$.
This function has been used as an alternative method for applying scattering boundary conditions
for one-electron problems~\cite{Fre90}, yielding similar results to the conventional boundary conditions. 
For one electron in one dimension, 
the real function $f(q,p)$ is defined by taking the Wigner transform of the density matrix 
$\rho(x,x^\prime)=\psi^*(x^\prime)\psi(x)$,
given by
\begin{equation}
f(q,p) = \int dr \, \rho(q+r/2,q-r/2)\, \exp(-ipr/\hbar),
\end{equation} 
and it has the advantage of echoing the properties of a classical distribution function
(aside for the requirement of strictly positive probabilities)
with $q$ and $p$ playing the r\^{o}le of position and momentum, 
thus giving a phase-space picture of quantum mechanics. 
For example, expectation values of operator functions $A(\hat{q})$ of position $\hat{q}$ may be expressed as
\begin{equation} 
<{\cal A}(\hat{q}) > = \frac{1}{2 \pi \hbar} \int \, dp \, dq \, {\cal A}(q) \, f(q,p)
\end{equation}
and similarly for functions of momentum $\hat{p}$.
The Wigner function also satisfies a quantum Liouville equation with a well-defined
classical limit as $\hbar \rightarrow 0$ leading to the Boltzmann transport equation.
Single-electron scattering boundary conditions can then be applied to a domain $0 \leq x \leq L$
by requiring that $f(0,p>0)$ and $f(L,p<0)$ be constrained to their equilibrium values. 
That is, the momentum distribution of the {\it incoming electrons} on the left and
the right is fixed to values characteristic of the reservoir.
The \Schro equation is then solved, subject to these boundary conditions,
thereby allowing the device to find a steady state by varying the values
of the Wigner function in the interior of the device region and at $f(0,p<0)$ and $f(L,p>0)$
(the {\it outgoing electrons}). We note that these boundary conditions break 
time-reversal symmetry as is necessary to generate a current-carrying state.

To find a form of the scattering boundary conditions applicable to the many-body problem,
we note that the same Wigner transform can be applied to the one-particle reduced density
matrix calculated from a {\it many-body wavefunction} $\Psi$, 
\begin{eqnarray}
\rho({\bf r},{\bf r^\prime}) = 
\sum_{s_1}		\int d \bx_2 \ldots d \bx_N && \! \! \! \! \! \! \! \!
	                   \Psi^*({\bf r^\prime} s_1, \bx_2, \ldots, \bx_N) \nonumber \\
\times && \! \! \! \! \! \! \! \Psi  \; \: (\br s_1, \bx_2, \ldots, \bx_N)
\end{eqnarray}
where $\bx_n = \br_n s_n$.
This generates the Wigner function $f(\bq,\bp)$ of the correlated wavefunction $\Psi$.
To simplify matters, we additionally integrate $f$ over planes perpendicular to the
current flow, yielding a one-dimensional Wigner function $f(q,p)$ as above, 
but our approach does not depend on this simplification.
Our example simulation uses the Au$_{13}$-benzene-dithiolate-Au$_{13}$ contact-molecule-contact system
to model the transport region, shown in fig. 1 (top).
\begin{figure}[t]
\includegraphics[width=2.6 in]{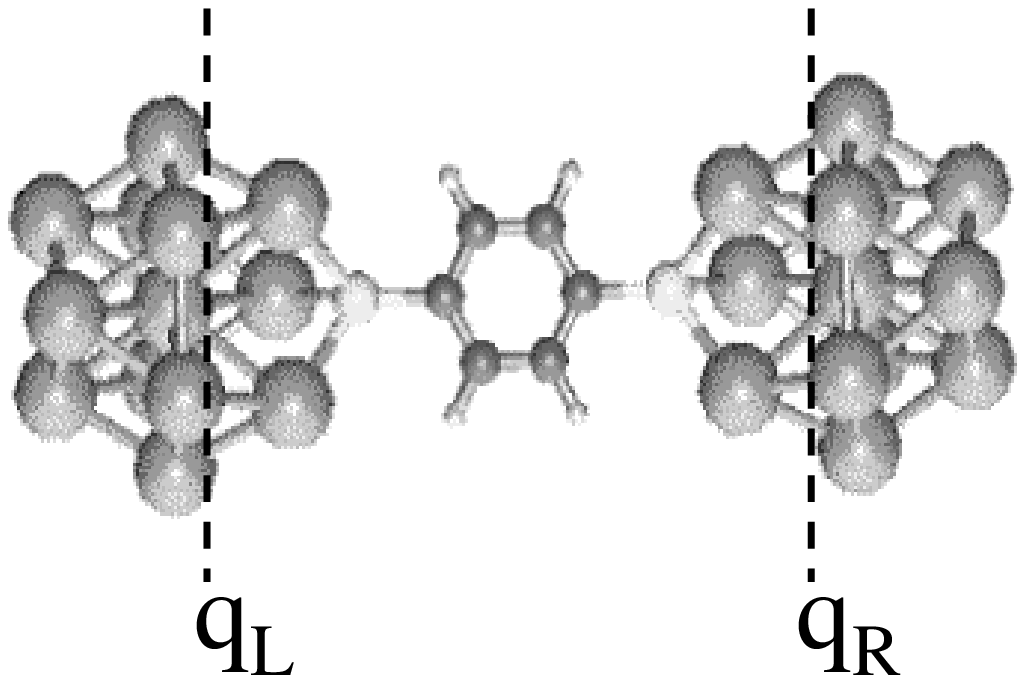}
\includegraphics[angle=-90.0,width=3.2 in ]{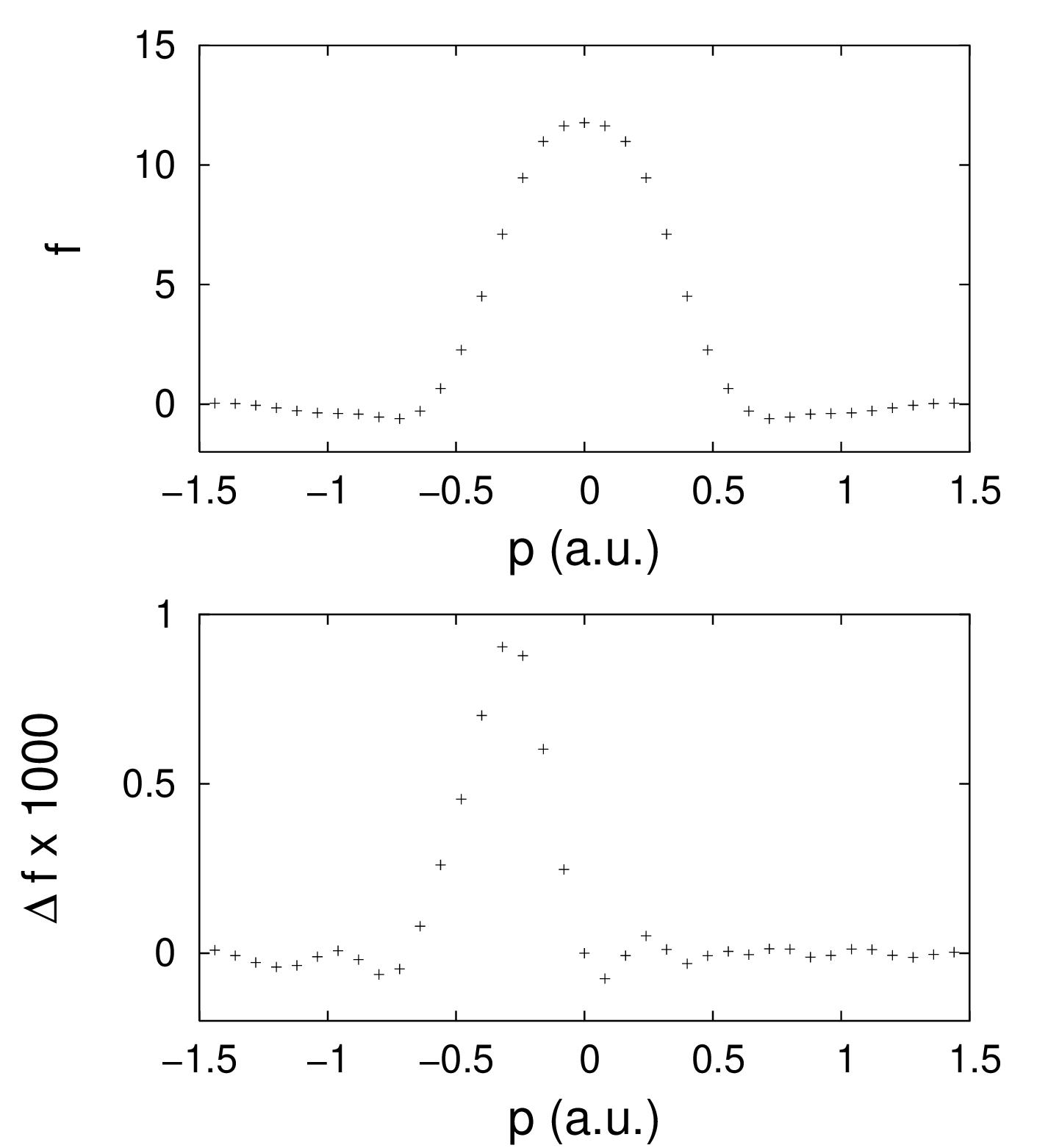}
\caption{({\it top}) Geometry of the Au$_{13}$-benzene-dithiol-Au$_{13}$ junction.
({\it middle}) The Wigner momentum distribution function at the left reservoir/device interface at
equilibrium ($V=0$). We use Hartree atomic units for $p$ ($m=1$,$c \simeq 137$).
({\it bottom}) The difference between the finite- and zero-bias Wigner functions at the left contact: note the change in scale.
The constraints have fixed the momentum distribution for momenta $p>0$, while allowing the
distribution at $p<0$ to vary.
}
\end{figure}
Having found the CI ground state of the transport region in the absence of voltage bias, 
we calculate the equilibrium values of the Wigner function in the left
and right contacts $f_0(q_L,p)$ for $p>0$ and $f_0(q_R,p)$ for $p<0$; 
the left contact values are shown in fig. 1 (middle).
We then formulate the quantum many-body transport problem by making the following {\it Ansatz}: 
the appropriate many-body wave function $\Psi$ for the transport region in the
presence of an applied electric field ${\cal E} \hat{{\rm z}}$ along the molecular axis minimizes the exact energy 
$ E=<\Psi|\hat{{\rm H}}+e {\cal E} \sum_n {\hat{\rm z}}_n |\Psi>$  
while simultaneously satisfying the open system boundary conditions
\begin{equation}
<\Psi| \hat{{\rm F}}(q_L,p) | \Psi> = f_0(q_L,p) \; \;  \mbox{for  } p>0 \label{eq:FwbcL}
\end{equation}
and 
\begin{equation}
<\Psi| \hat{{\rm F}}(q_R,p) | \Psi>  = f_0(q_R,p) \; \; \mbox{for  } p<0 \label{eq:FwbcR}
\end{equation}
with normalisation constraint 
$<\!\!\!\!\Psi|\Psi\!\!\!\!> = 1$ and constraints to enforce current continuity.
Here $\hat{{\rm F}}(q,p)$ is the hermitian operator corresponding to the quantity $f(q,p)$.
Formally, our {\it Ansatz} corresponds to the
zero-temperature maximum entropy principle on the transport region with the boundary conditions specified as
system observables \cite{Jay57}.

To numerically implement this scheme, we replace the infinite
number of constraints implicit in equations~(\ref{eq:FwbcL}),(\ref{eq:FwbcR}) with a finite set of constraints
on a grid in momentum space. 
To solve this {\it non-linear constrained minimization problem}
we introduce Lagrange multipliers $\lambda_i$ for each constraint,
giving us a simpler {\it unconstrained} minimization problem of the associated Lagrangian function
\begin{equation}
L=<\Psi | \hat{{\rm H}} + e{\cal E} \sum_n \hat{{\rm z}}_n | \Psi> - \sum_i \lambda_i <\Psi | \hat{{\rm C}}^i | \Psi> \label{eq:LF}
\end{equation}
where the $\hat{{\rm C}}^i$ are the hermitian operators corresponding to the constraint observables.
We need to determine the correct values of the Lagrange multipliers;
these are not known {\it a priori}, 
but can be found by an iterative optimization technique~\cite{NoW99}.
In figure 1 (bottom) we show the difference between a finite bias and equilibrium Wigner distribution and verify
that the minimization procedure respects the boundary constraints.
Electron flow results from the difference between these distributions,
giving a net electric current to the right. 
Once the minimising wavefunction $\Psi$ has been obtained, 
we find the current $I$ flowing simply by integrating the probability current density ${\bf J} (\br)$
over planes perpendicular to the molecular axis.

For our prototype transport calculation, 
we begin with a fully relaxed structure (no strain at zero current) consisting of the benzene-dithiolate
molecule bonded between the (111) faces of two gold 
clusters in $C_{2v}$ symmetry \cite{LNG02}; see fig. 1 (top).
A set of many-body expansion functions for this structure 
was selected using the Monte Carlo configuration interaction method \cite{Gre98,ci_details}.
Exact spin coupling is applied and the zero-bias total system eigenfunction $\Psi_0$ is a singlet for the ground state.
In fig. 2, we present our computed current-voltage (IV) characteristics for the benzene-dithiol system. 
The IV curve has been generated by increasing the applied field ${\cal E}$ in steps and finding the
energy-minimising wavefunction at each field. 
The voltage is then calculated by integrating the external electric field
between the two reservoir/device interface planes where the Wigner constraints are applied. 

\begin{figure}[ht]
\includegraphics[angle=-90.0,width=3.2 in]{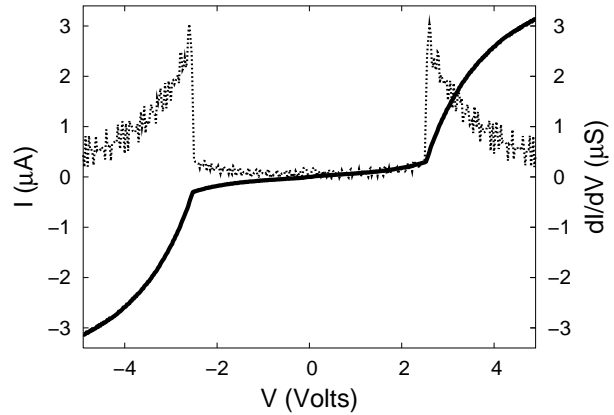}
\caption[]{Calculated current-voltage characteristics for benzene-dithiolate bonded 
to the (111) faces of two gold clusters; 
also shown is the differential conductance $\frac{d I}{d V}$.
}
\end{figure}

We find two distinct regions in the resulting IV characteristics,
with an abrupt transition between the two near 2.5 V.
We can trace the origins of this transition by examining the polarization behaviour
of the {\it isolated} Au$_{13}$-benzene-dithiolate-Au$_{13}$ system under application of an
external field.
By removing the Wigner constraints we decouple the transport region from the gold wires;
in fig.\ 3 we show the energies of the first five eigenstates as the applied field is increased.
The energy levels are labelled by the irreducible representation of $C_{2v}$ by
which they transform at zero field, and by energy ordering within this irrep. 
The system ground state belongs to $A_1$, and at finite field mixes with states of $B_2$ symmetry.
Second order perturbation theory predicts that 
the ground state and first excited state $|1B_2>$ will respond slowly to the applied field, 
as they are isolated in energy and so have large energy denominators, as we indeed see.
In contrast, the three states $|2A_1\!\!>, |3A_1\!\!>$ and $|2B_2\!\!>$ can
couple strongly, and a state associated with these excitations
polarizes rapidly with the external electric field, and crosses the ground state at approximately 2.5 V.
Furthermore, the rigorous definition of correlation involves physics beyond that describable by a single Slater determinant.
The eigenvalues of the reduced density matrix $\rho({\bf r},{\bf r^\prime})$
are the {\it natural orbital occupancies} $n_i$;
a necessary condition for $\Psi$ to originate from some single determinant is that $n_i=2$ for $N/2$ occupied orbitals
and $n_i=0$ for all others.
The CI wavefunction at $V=0$ has a typical pattern of occupancies slightly changed from that of a single determinant,
and this remains the same up to the transition. 
Above the transition, however, we see a significant number of orbitals whose occupancies differ from 2 or 0.
Taken together with the excited state influence on the transition voltage, 
we see that electron correlation is essential for an accurate description of molecular electronic transport.

\begin{figure}
\includegraphics[angle=-90.0,width=3.2in]{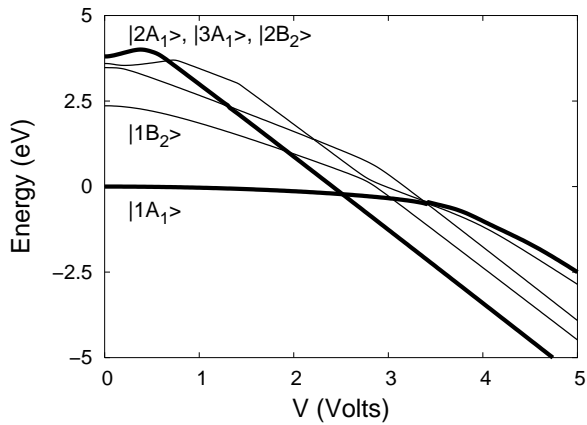}
\caption{Energies of the lowest five CI eigenstates as a function of applied field,
for the isolated (no Wigner constraints) Au$_{13}$-benzene-dithiolate-Au$_{13}$ cluster. 
}
\end{figure}

We note here that the current magnitude below resonance is in good agreement with experiment \cite{RZM97},
particularly when compared to other theoretical predictions, and the qualitative behaviour of the IV characteristic
is consistent with other recent findings of molecular scale transport, with non-resonant tunnel regimes
followed by a rapid increase in current magnitude~\cite{Lee03}. 
Our calculations indicate an onset resistance
of 18 M$\Omega$ and a resonant tunnel resistance of several M$\Omega$,
close to the corresponding experimental values of 
$\approx$ 22 M$\Omega$ and $\approx 13$ M$\Omega$ estimated by Reed {\it et al}.
Previous {\it ab initio} DFT calculations typically find currents in the range
of $20\ 000$ to $60\ 000$ nAmps at 2 V \cite{DPL00,STB03},
whereas we find a current of 160 nAmps which compares
favourably with the experimental measurement of $\approx 60$ nAmps~\cite{RZM97}.

In conclusion, we have presented an approach to electronic transport that represents
a radical departure from either scattering theory or Green's function methods. 
The formalism is conceptually simple, 
and is made possible by a generalisation of the scattering boundary conditions to many-body calculations.
Our method yields many-body wavefunctions which are exact eigenstates of spin
and so can also be used for studying spin-dependent transport and Kondo effect physics;
we can also easily add a gate field to the Hamiltonian.
For the Au$_{13}$-benzene-dithiolate-Au$_{13}$ system, we calculate
electronic currents orders of magnitude lower than other theoretical predictions, but of the correct
order of magnitude with respect to the experimental measurements, the first such result to our knowledge.
We hypothesize that as DFT exchange correlation functionals have been developed
to reproduce an integrated quantity (the energy), 
they must be modified for accurate treatment of transport 
by determining exchange-correlation potentials that are locally accurate.
We note that for accurate molecular electronics simulation, it appears that a highly correlated 
treatment of the electronic structure is needed. 

{\bf Acknowledgments} This work was supported by the European Union's Future \& Emerging Technology programme
through the Nanotcad project and by Science Foundation Ireland.


\begin{thebibliography}{99}

\bibitem{RZM97}
M.A. Reed {\it et al.}, Science {\bf 278}, 252 (1997).

\bibitem{JGA00}
C. Joachim, J.K. Gimzewski and A. Aviram, Nature {\bf 408}, 541 (2000).

\bibitem{NiR03}
A. Nitzan and M.A. Ratner, Science {\bf 300}, 1384 (2003).

\bibitem{EmK98}
E.G. Emberly and G. Kirczenow, Phys. Rev. B {\bf 58}, 10911 (1998).

\bibitem{DPL00}
M. Di Ventra, S.T. Pantelides and N.D. Lang, Phys. Rev. Lett. {\bf 84}, 979 (2000).

\bibitem{DeS01}
P.A. Derosa and J.M. Seminario, J. Phys. Chem. B {\bf 105}, 471 (2001).

\bibitem{DGD01}
P.S. Damle, A.W. Ghosh and S. Datta, Phys. Rev. B {\bf 64}, 201403 (2001).

\bibitem{STB03}
K. Stokbro {\it et al.}, Comp. Mat. Sci. {\bf 27} 151, (2003).

\bibitem{Krs03}
P.S. Krsti\'c {\it et al.}, Comp. Mat. Sci. {\bf 28}, 321 (2003).

\bibitem{LanBue}
R. Landuaer, IBM. J. Res. Develop. {\bf 1}, 233 (1957);
M. B\"uttiker, Y. Imry, R. Landauer and S. Pinhas, Phys. Rev. B {\bf 31}, 6207 (1985).

\bibitem{Lee03}
J.O. Lee {\it et al.}, Nanoletters {\bf 3}, 113 (2003).

\bibitem{ViR87}
G. Vignale and M. Rasolt, Phys. Rev. Lett. {\bf 59}, 2360 (1987).

\bibitem{Fre90}
W. R. Frensley, Rev. Mod. Phys. {\bf 62}, 745 (1990).

\bibitem{Jay57}
E.T. Jaynes, Phys. Rev. {\bf 106}, 620 (1957).

\bibitem{NoW99}
J. Nocedal and S.J. Wright, Numerical Optimization, Springer Verlag (1999).

\bibitem{LNG02}
J.A. Larsson, M. Nolan and J.C. Greer, J. Phys. Chem. B {\bf 106}, 5931 (2002).

\bibitem{Gre98}
J.C. Greer, J. Comp. Phys. {\bf 146}, 181 (1998).

\bibitem{ci_details}
Effective core potentials were applied to all
gold atoms (78 core electrons) and to the bonding sulfur atoms (10 core electrons). 
Atomic basis sets: Au, $sp$ on 20 atoms, $spd$ on the 6 bonding gold atoms;
S, $2s2p$; C $4s2p1d$; H $2s$.
Many-body basis set: the three lowest zero-field $A_1$ and $B_2$ CI eigenstates were determined,
using an acceptance criteria $c_{\rm min}=10^{-2}$.
We concatenated these six configuration sets to form the transport basis set.

\end{thebibliography}
\end{document}